\documentclass{article}
\usepackage{cite}
\usepackage{graphicx}
\usepackage{dcolumn}
\usepackage{amsmath}

\begin{document}

\date{}
\title{On some conditionally solvable quantum-mechanical problems}
\author{Paolo Amore\thanks{%
e--mail: paolo@ucol.mx} \\
Facultad de Ciencias, CUICBAS, Universidad de Colima,\\
Bernal D\'{\i}az del Castillo 340, Colima, Colima,Mexico \\
and \\
Francisco M. Fern\'andez\thanks{%
e--mail: fernande@quimica.unlp.edu.ar} \\
INIFTA, Divisi\'{o}n Qu\'{\i}mica Te\'{o}rica,\\
Blvd. 113 y 64 (S/N), Sucursal 4, Casilla de Correo 16,\\
1900 La Plata, Argentina}
\maketitle

\begin{abstract}
We analyze two conditionally solvable quantum-mechanical models: a
one-dimensional sextic oscillator and a perturbed Coulomb problem. Both lead
to a three-term recurrence relation for the expansion coefficients. We show
diagrams of the distribution of their exact eigenvalues with the addition of
accurate ones from variational calculations. We discuss the symmetry of such
distributions. We also comment on the wrong interpretation of the exact
eigenvalues and eigenfunctions by some researchers that has led to the
prediction of allowed cyclotron frequencies and field intensities.
\end{abstract}

\section{Introduction}

\label{sec:intro}

In addition to the exactly solvable quantum-mechanical models, like the
harmonic oscillator and the hydrogen atom, among others, where one obtains
the whole spectrum for any values of the model parameters\cite{LL65,CDL77},
there is the class of quasi-solvable or conditionally-solvable systems where
one obtains a subset of eigenvalues and eigenfunctions in exact analytical
way only for some values of the model parameters or suitable constrains for
them\cite
{F79,FD80,M81,F81,F81b,RV91,FW81,D88,K89,BV90,CM95,CDW00,D02,IS07,BCD17}.
The most widely studied models are anharmonic oscillators\cite
{F79,FD80,M81,F81,F81b,FW81,K89,BV90,CM95,CDW00,D02,IS07} and perturbed
Coulomb systems\cite{D88,CM95,D02,IS07,BCD17}.

The perturbed Coulomb system appears in the analysis of a variety of
physical problems and enabled some researchers to predict the existence of
allowed cyclotron frequencies, allowed field intensities, etc\cite
{V91,MHV92,FDMBB94,BM12,BB12,B14,B14b,BB14,FB15,BF15,FB12,BB13b,HMM20}.

The purpose of this paper is the analysis of two conditionally solvable
models that can be reduced to a three-term recurrence relation. In Sec.~\ref
{sec:Rec_rel} we outline the main features of the three-term recurrence
relation, in sections \ref{sec:Example 1} and \ref{sec:Example 2} we show
diagrams for the distribution of the eigenvalues of a sextic anharmonic
oscillator and a perturbed Coulomb model, respectively. We comment on the
wrong interpretation of the eigenvalues obtained by several authors for the
latter example. Finally, in Sec.~\ref{sec:conclusions} we summarize the main
results and draw conclusions.

\section{The recurrence relation}

\label{sec:Rec_rel}

Consider a Schr\"{o}dinger equation
\begin{equation}
H\psi =E\psi ,  \label{eq:Schro}
\end{equation}
with a Hamiltonian operator $H=H(\mathbf{a})$ that depends on a set of model
parameters $\mathbf{a}=\left( a_{1},a_{2},\ldots a_{K}\right) $. Suppose
that we can write the solution $\psi $ as a linear combination of a (not
necessarily orthonormal) basis set $\left\{ \varphi _{j},\,j=0,1,\ldots
\right\} $%
\begin{equation}
\psi =\sum_{j=0}^{\infty }c_{j}\varphi _{j},  \label{eq:psi_expansion}
\end{equation}
so that the coefficients $c_{j}$ obey a three-term recurrence relation
\begin{equation}
c_{j+1}=A_{j}c_{j}+B_{j}c_{j-1},\,j=0,1,\ldots ,\,c_{-1}=0,\,c_{0}=1.
\label{eq:three-term-rec-rel}
\end{equation}
Some authors state that the expansion coefficients are
normalization constants\cite{D02,IS07}.

If the equations
\begin{equation}
c_{n+1}=c_{n+2}=0,\,c_{n}\neq 0,  \label{eq:trunc_cond_1}
\end{equation}
can be solved for $E$ and $\mathbf{a}$ for some $n$, then $c_{j}=0$ for all $%
j>n$. If the solutions to these equations are physically acceptable then we
have obtained an exact solution to the Schr\"{o}dinger equation (\ref
{eq:Schro}). More precisely, we would have obtained $E=E^{(n)}(\mathbf{a}%
_{n})$, where the set of parameters $\mathbf{a}_{n}$ is a solution to some
nonlinear equation $F_{n}\left( \mathbf{a}_{n}\right) =0$. Clearly, the
model parameters $\mathbf{a}_{n}$ that satisfy this condition will not be
independent. The truncation condition (\ref{eq:trunc_cond_1}) is equivalent
to
\begin{equation}
B_{n+1}=0,\,c_{n+1}=0,\,c_{n}\neq 0.  \label{eq:trunc_cond_2}
\end{equation}
This truncation condition proposed by Ver\c{c}in\cite{V91} for a particular
problem appears to be simpler than the determinantal condition used by other
authors\cite{BV90,CDW00}.

The coefficients $\hat{c}_{j}=(-1)^{j}c_{j}$ satisfy
\begin{equation}
\hat{c}_{j+1}=-A_{j}\hat{c}_{j}+B_{j}\hat{c}_{j-1},\,j=0,1,\ldots ,\,\hat{c}%
_{-1}=0,\,\hat{c}_{0}=1,  \label{eq:TTRR2}
\end{equation}
that will be useful for the interpretation of some of the results below.

\section{One-dimensional anharmonic oscillator}

\label{sec:Example 1}

As a first example we consider the anharmonic oscillator
\begin{eqnarray}
H &=&-\frac{d^{2}}{dx^{2}}+V(a,b,x),\,  \nonumber \\
V(a,b,x) &=&-ax^{2}-bx^{4}+x^{6},\,-\infty <x<\infty ,  \label{eq:H_x2x4x6}
\end{eqnarray}
that supports bound states for all real values of the parameters $a$ and $b$%
. In earlier treatments of the sextic oscillator the authors considered a
positive coefficient for the sextic term\cite
{F79,FD80,M81,F81,F81b,FW81,K89,BV90,CM95,CDW00,D02,IS07}. However, such a
coefficient can be easily set to unity by means of a suitable change of the
independent variable $x$\cite{F20}. For this reason we choose the
coefficient of $x^{6}$ equal to unity without loss of generality.

We have three cases:

Case I: $4a+b^{2}<0\Rightarrow $ single well

Case II: $4a+b^{2}>0,\,a>0\Rightarrow $ double well

Case III: $4a+b^{2}>0,\,a<0\Rightarrow $ triple well

A suitable non-orthogonal basis set for the treatment of this problem is
\begin{equation}
\varphi _{j,s}(x)=x^{s+2j}\exp \left( \frac{b}{4}x^{2}-\frac{1}{4}%
x^{4}\right) ,\,j=0,1,\ldots ,  \label{eq:phi_x2x4x6}
\end{equation}
where $s=0$ or $s=1$ for even or odd states, respectively. A straightforward
calculaton leads to
\begin{eqnarray}
A_{j} &=&-\frac{\left( b\left( 4j+2s+1\right) +2E\right) }{4\left(
j+1\right) \left( 2j+2s+1\right) },  \nonumber \\
B_{j} &=&-\frac{\left( 4a+b^{2}-4\left( 4j+2s-1\right) \right) }{8\left(
j+1\right) \left( 2j+2s+1\right) }.  \label{eq:A_j_B_j_x2x4x6}
\end{eqnarray}
From $B_{n+1}=0$ we obtain a relationship for the model parameters
\begin{equation}
4a+b^{2}-4\left( 4n+2s+3\right) =0,  \label{eq:F(a,b)=0_x2x4x6}
\end{equation}
from which we obtain either $a(b)$ or $b(a)$. Notice that the truncation
condition gives us the possibility of double and triple wells.

The truncation condition (\ref{eq:trunc_cond_2}) leads to an exact
eigenfunction
\begin{eqnarray}
\psi _{s}^{(n)}(x) &=&x^{s}P_{s}^{(n)}(x)\exp \left( \frac{b}{4}x^{2}-\frac{1%
}{4}x^{6}\right) ,  \nonumber \\
P_{s}^{(n)}(x) &=&\sum_{j=0}^{n}c_{s,j}^{(n)}x^{2j}.  \label{eq:psi_x2x4x6}
\end{eqnarray}
If we solve equation (\ref{eq:F(a,b)=0_x2x4x6}) for $a$ we obtain $%
a_{n,s}(b) $ and a Hamiltonian operator $H_{n,s}$ that depends on this
particular relationship between $a$ and $b$. It is worth having in mind that
$\psi _{s}^{(n)}(x)$ and $\psi _{s^{\prime }}^{(n^{\prime })}(x)$ are not
eigenfunctions of the same hamiltonian but of $H_{n,s}$ and $H_{n^{\prime
},s^{\prime }}$, respectively, which are two different operators. We are not
being unnecessarily careful about this point because it has been
misunderstood by other authors\cite
{V91,MHV92,FDMBB94,BM12,BB12,B14,B14b,BB14,FB15,BF15,FB12,BB13b,HMM20}.

For example, for $n=0$ we obtain
\begin{equation}
a_{0,s}(b)=2s+3-\frac{b^{2}}{4},\,E_{s}^{(0)}=-\frac{b\left( 2s+1\right) }{2}%
.
\end{equation}
Since $c_{s,j}^{(0)}=0$ for all $j>0$ the corresponding function has no
nodes when $s=0$ and only one node at $x=0$ when $s=1$. The former is the
ground state for the model determined by $a_{0,0}(b)$ and the latter is the
first-excited state for $a_{0,1}(b)$.

For $n=1$ we have two solutions
\begin{eqnarray}
a_{1,s} &=&2s+7-\frac{b^{2}}{4},  \nonumber \\
E_{0,s}^{(1)} &=&-\frac{b\left( 2s+3\right) +2\sqrt{b^{2}+8\left(
2s+1\right) }}{2},\,  \nonumber \\
E_{1,s}^{(2)} &=&-\frac{b\left( 2s+3\right) -2\sqrt{b^{2}+8\left(
2s+1\right) }}{2}.
\end{eqnarray}
The values of the coefficient $c_{1}$ in these two cases are
\begin{equation}
c_{0,s,1}^{(1)}=\frac{b+\sqrt{b^{2}+8\left( 2s+1\right) }}{2\left(
2s+1\right) },\,c_{1,s,1}^{(2)}=\frac{b-\sqrt{b^{2}+8\left( 2s+1\right) }}{%
2\left( 2s+1\right) }.
\end{equation}
Since $c_{0,s,1}^{(1)}>0$ and $c_{1,s,1}^{(2)}<0$ we conclude that they are
consistent with the first two even states of $H_{1,0}$ and the fist two odd
states of $H_{1,1}$.

For $n=2$ we have three eigenvalues for just one value of $a$
\begin{eqnarray}
a_{2,s} &=&2s+11-\frac{b^{2}}{4},  \nonumber \\
8E^{3} &+&12b\left( 2s+5\right) E^{2}+2\left[ b^{2}\left(
12s^{2}+60s+59\right) -256\left( s+1\right) \right] E  \nonumber \\
&+&b\left( 2s+1\right) \left[ b^{2}\left( 2s+5\right) \left( 2s+9\right)
-256\left( s+3\right) \right] =0
\end{eqnarray}
The three roots $E_{0,s}^{(2)}<E_{1,s}^{(2)}<E_{2,s}^{(2)}$ are three
eigenvalues of the anharmonic oscillator $H_{2,s}$, even states for $s=0$
and odd ones for $s=1$. In general we expect $n+1$ energy eigenvalues $%
E_{i,s}^{(n)}$, $i=0,1,\ldots ,n$ for the model given by the curve $%
a_{n,s}(b)$. In Appendix~\ref{sec:Symmetric matrix} we prove that all the
roots $E_{i,s}^{(n)}$ are real. Notice that we have chosen the subscripts so
that the eigenfunction $\psi _{i,s}^{(n)}(x)$ has exactly $i+s$ nodes and we
can consider both $i$ and $s$ to be quantum numbers. In other words, we can
label the eigenfunctions and eigenvalues of the Hamiltonian operator (\ref
{eq:H_x2x4x6}) as $\psi _{i,s}(x)$ and $E_{i,s}(a,b)$, respectively. On the
other hand, the integer $n$ cannot be considered a quantum number as it
merely indicates a given relationship $F_{n,s}(a,b)=0$ between the
parameters $a$ and $b$ from which we obtain either $a_{n,s}(b)$ or $%
b_{n,s}(a)$ and the Hamiltonian operator $H_{n,s}$. Although the truncation
condition (\ref{eq:trunc_cond_2}) leads to a Hamiltonian operator that
depends on $s$, this integer is actually a quantum number associated to the
parity of the eigenfunction. For example, the well known eigenvalues $E_{\nu
}=\hbar \omega \left( \nu +\frac{1}{2}\right) $, $\nu =0,1,\ldots $ of the
harmonic oscillator can be written as $E_{i,s}=\hbar \omega \left( 2i+s+%
\frac{1}{2}\right) $, $i=0,1,\ldots $, $s=0,1$.

\begin{figure}[tbp]
\begin{center}
\includegraphics[width=9cm]{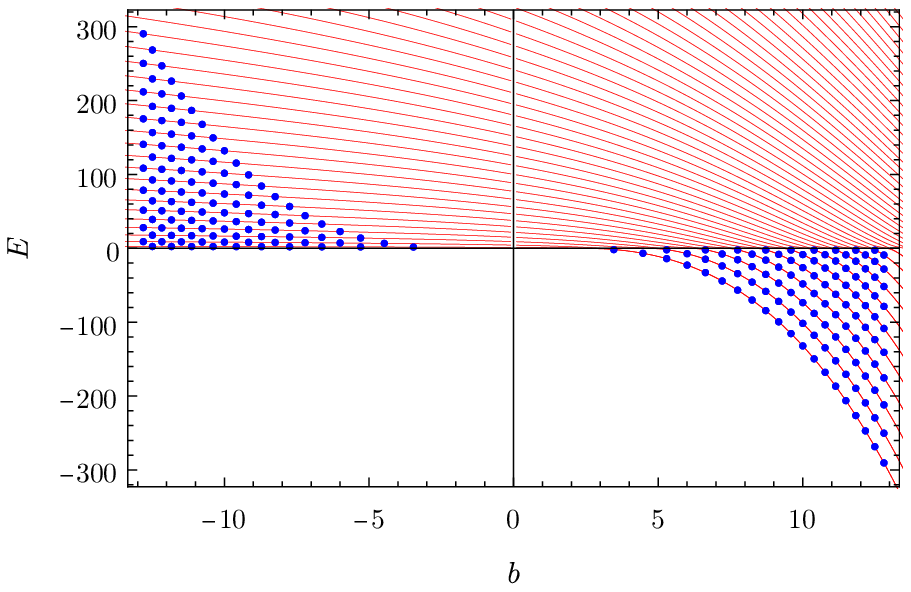}
\end{center}
\caption{Eigenvalues of the one-dimensional anharmonic oscillator $%
E_{\nu,s}(0,b)$ calculated by means of the truncation condition (blue
points) and by a variational method (red lines) }
\label{fig:AHOa0}
\end{figure}

Figure~\ref{fig:AHOa0} shows the eigenvalues $E_{\nu ,s}(0,b)$ obtained from
the truncation condition (blue circles) and those calculated by means of two
variational methods\cite{O87,A06} (red lines). The potential-energy function
$V(0,b,x)$ is a single well for $b<0$ and a double well for $b>0$. Since the
depth of the wells increases with $b$ one expects negative eigenvalues for
sufficiently large values of $b$. Straightforward calculation using the
Riccati-Pad\'{e} method (RPM)\cite{FMT89a} shows that the first eigenvalue
becomes negative at $b=2.491322600$ and the second one at $b=3.037089563$.
There is a gap without blue points because the truncation condition requires
that $b^{2}\geq 12$. Notice the coalescence of pairs of even and odd states
as the wells become deeper. The members of such pairs approach each other
when $b$ increases.

The symmetry of Figure~\ref{fig:AHOa0} can be easily explained by an
argument similar to that given by Child et al\cite{CDW00} for the
central-field version of this model. To this end, notice that $%
A_{j}(-b,-E)=-A_{j}(b,E)$ and $B_{j}(a,-b)=B_{j}(a,b)$ leads to equation (%
\ref{eq:TTRR2}) with solutions $\hat{c}_{j}$. It is worth mentioning that
the roots of the Hankel-Hadamard determinants in the Riccati-Pad\'{e} method
(RPM)\cite{FMT89a} yield the eigenvalues for $b>0$ and $b<0$ simultaneously.

\begin{figure}[tbp]
\begin{center}
\includegraphics[width=9cm]{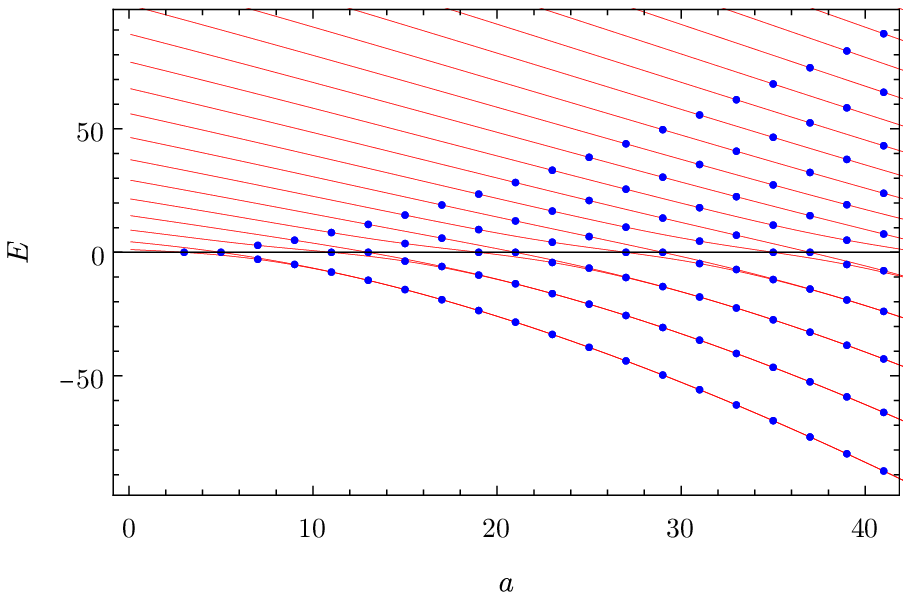}
\end{center}
\caption{Eigenvalues $E_{\nu,0s}(a,0)$ of the one-dimensional anharmonic
oscillator calculated by means of the truncation condition (blue points) and
by a variational method (red lines) }
\label{fig:AHOb0}
\end{figure}

Figure~\ref{fig:AHOb0} shows eigenvalues $E_{\nu ,s}(a,0)$ obtained in the
same way. The behaviour of these eigenvalues is similar to those in the
previous case, except for the lack of symmetry. In this case the first
eigenvalue becomes negative at $a=3$ and the second one at $a=5$.

\begin{figure}[tbp]
\begin{center}
\includegraphics[width=9cm]{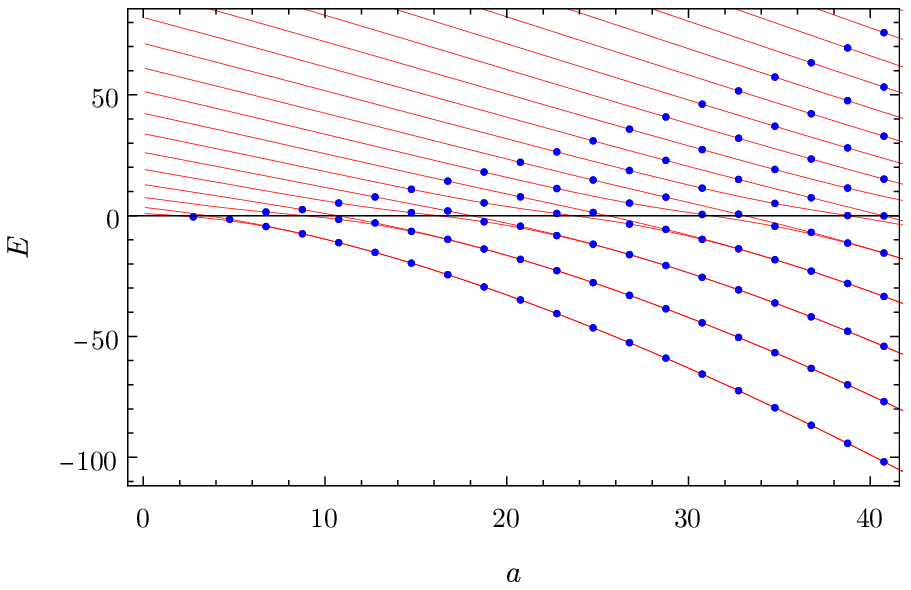}
\end{center}
\caption{Eigenvalues $E_{\nu,s}(a,1)$ of the one-dimensional anharmonic
oscillator calculated by means of the truncation condition (blue points) and
a variational method (red lines) }
\label{fig:AHOb1}
\end{figure}

Figure~\ref{fig:AHOb1} shows eigenvalues $E_{\nu ,s}(a,1)$. The first
eigenvalue becomes negative at $a=1.901043863$ and the second one at $%
a=3.508348408$.

The behaviour of the eigenvalues $E_{\nu ,s}(a,b)$ with respect to the model
parameters $a$ and $b$ is given by the celebrated Hellmann-Feynman theorem%
\cite{P68} (and references therein)
\begin{equation}
\frac{\partial E}{\partial a}=-\left\langle x^{2}\right\rangle ,\,\frac{%
\partial E}{\partial b}=-\left\langle x^{4}\right\rangle .
\label{eq:HFTx2x4x6}
\end{equation}

\section{Perturbed Coulomb model}

\label{sec:Example 2}

The second example is given by the Hamiltonian operator
\begin{eqnarray}
H &=&-\frac{d^{2}}{dr^{2}}+\frac{\gamma (\gamma +1)}{r^{2}}+V(a,b,r),\,
\nonumber \\
V(a,b,r) &=&-\frac{a}{r}-br+r^{2},\,0\leq r<\infty ,  \label{eq:H_CLH}
\end{eqnarray}
where $\gamma >0$ and $a$ and $b$ are real. Notice that this form of the
Hamiltonian operator is suitable for the treatment of the central field
model in any number of spatial dimensions. In fact, $\gamma $ may be a
function of the number of spatial dimensions $D$ and the rotational quantum
number $l$\cite{CM95,D02,IS07,BCD17} and may even take into account a term
in the potential-energy function that behaves as $r^{-2}$ at origin\cite{D88}%
.

A suitable basis set for this problem is
\begin{equation}
\varphi _{j}(r)=r^{\gamma +1+j}\exp \left[ \frac{b}{2}r-\frac{r^{2}}{2}%
\right] ,\,j=0,1,2,\ldots ,  \label{eq:phi_CLH}
\end{equation}
and we obtain the three-term recurrence relation (\ref{eq:three-term-rec-rel}%
) with
\begin{eqnarray}
A_{j} &=&-\frac{a+b\left( j+\gamma +1\right) }{\left( j+1\right) \left[
j+2\left( \gamma +1\right) \right] },  \nonumber \\
B_{j} &=&-\frac{b^{2}+4\left( E-2j-2\gamma -1\right) }{4\left( j+1\right)
\left[ j+2\left( \gamma +1\right) \right] }.  \label{eq:A_j_B_j_CLH}
\end{eqnarray}
From $B_{n+1}=0$ we obtain an expression for the energy
\begin{equation}
E_{\gamma }^{(n)}=2\gamma +2n+3-\frac{b^{2}}{4},
\end{equation}
and the truncation condition (\ref{eq:trunc_cond_2}) leads to wavefunctions
of the form
\begin{eqnarray}
\psi _{\gamma }^{(n)} &=&r^{\gamma +1}P_{\gamma }^{(n)}(r)\exp \left[ \frac{b%
}{2}r-\frac{r^{2}}{2}\right] ,  \nonumber \\
P_{\gamma }^{(n)}(r) &=&\sum_{j=0}^{n}c_{j}r^{j}.  \label{eq:psi_CLH}
\end{eqnarray}
For simplicity, we label both the eigenvalues and eigenfunctions with the
real number $\gamma $ although in general it will not be a true quantum
number. We follow this practice because the form of $\gamma$ changes from
one model to another and it will commonly depend on the angular quantum
number\cite{D88,CM95,D02,IS07,BCD17}.

For $n=0$ we have
\begin{equation}
E_{\gamma }^{(0)}=2\gamma +3-\frac{b^{2}}{4},\,a_{0,\gamma }=-b\left( \gamma
+1\right) ,
\end{equation}
and the corresponding wavefunction $\psi _{\gamma }^{(0)}$ does not have
nodes.

For $n=1$ we have
\begin{eqnarray}
E_{\gamma }^{(1)} &=&2\gamma +5-\frac{b^{2}}{4},  \nonumber \\
a_{\gamma }^{(1,1)} &=&-\frac{b\left( 2\gamma +3\right) +\sqrt{%
b^{2}+16\left( \gamma +1\right) }}{2},  \nonumber \\
a_{\gamma }^{(1,2)} &=&-\frac{b\left( 2\gamma +3\right) -\sqrt{%
b^{2}+16\left( \gamma +1\right) }}{2}.
\end{eqnarray}
The wave function for $a_{\gamma }^{(1,1)}$ will not have nodes in the
interval $0<r<\infty $ because
\begin{equation}
c_{\gamma ,1}^{(1,1)}=\frac{\sqrt{b^{2}+16\left( \gamma +1\right) }+b}{%
4\left( \gamma +1\right) },
\end{equation}
is always positive. On the other hand, for the model $a_{\gamma }^{(1,2)}(b)$
we have one node in that interval because
\begin{equation}
c_{\gamma ,1}^{(1,2)}=\frac{b-\sqrt{b^{2}+16\left( \gamma +1\right) }}{%
4\left( \gamma +1\right) },
\end{equation}
is always negative.

For $n=2$ we have
\begin{eqnarray}
E_{\gamma }^{(2)} &=&2\gamma +7-\frac{b^{2}}{4},  \nonumber \\
a^{3} &+&3a^{2}b\left( \gamma +2\right) +a\left[ b^{2}\left( 3\gamma
^{2}+12\gamma +11\right) -4\left( 4\gamma +5\right) \right]  \nonumber \\
&+&b\left( \gamma +1\right) \left[ b^{2}\left( \gamma +2\right) \left(
\gamma +3\right) -4\left( 4\gamma +9\right) \right] =0.
\end{eqnarray}

In the general case we obtain $E_{\gamma }^{(n)}$ and $a_{n,\gamma
}^{(i)}(b) $, $i=1,2,\ldots ,n+1$, $a_{n,\gamma }^{(i+1)}(b)>a_{n,\gamma
}^{(i)}(b)$. In Appendix~\ref{sec:Symmetric matrix} we prove that all the
roots $a_{n,\gamma }^{(i)}(b)$ are real. In order to understand the
relationship between these results and the actual eigenvalues $E_{\nu
,\gamma }(a,b)$, $\nu =0,1,\ldots $ of the operator (\ref{eq:H_CLH}) we
resort to the Hellmann-Feynman theorem that in this case states that
\begin{equation}
\frac{\partial E}{\partial a}=-\left\langle \frac{1}{r}\right\rangle ,\,%
\frac{\partial E}{\partial b}=-\left\langle r\right\rangle .
\end{equation}
Since $E$ decreases with $a$ we conclude that the pair $\left[ E_{\gamma
}^{(n)},a_{n,\gamma }^{(i)}(b)\right] $ is a point on the curve $%
E_{i-1,\gamma }(a,b)$ for $a=a_{n,\gamma }^{(i)}(b)$.

It is clear that $\left[ E_{\gamma }^{(n)},\psi _{\gamma }^{(n)}\right] $ is
a pair of eigenvalue-eigenfunction of the operator $H_{n,\gamma }$, and that
$\left[ E_{\gamma ^{\prime }}^{(n^{\prime })},\psi _{\gamma ^{\prime
}}^{(n^{\prime })}\right] $ corresponds to $H_{n^{\prime },\gamma ^{\prime
}} $. This apparently obvious fact has been misunderstood in many papers and
the belief that $E_{\gamma }^{(n)}$ gives us the spectrum of a single
quantum-mechanical system has led to the wrong conclusion that there exist
allowed cyclotron frequencies, allowed field intensities and the like\cite
{V91,MHV92,FDMBB94,BM12,BB12,B14,B14b,BB14,FB15,BF15,FB12,BB13b,HMM20}. Such
wrong conjectures arise from the belief that there are no square-integrable
solutions outside those given by the truncation condition (\ref
{eq:trunc_cond_2}). This misinterpretation of the meaning of the exact
solutions to conditionally solvable models has led to the fictitious
dependence of frequencies and field intensities on the quantum numbers
through, for instance, the parameter $a_{n,\gamma }$.

\begin{figure}[tbp]
\begin{center}
\includegraphics[width=9cm]{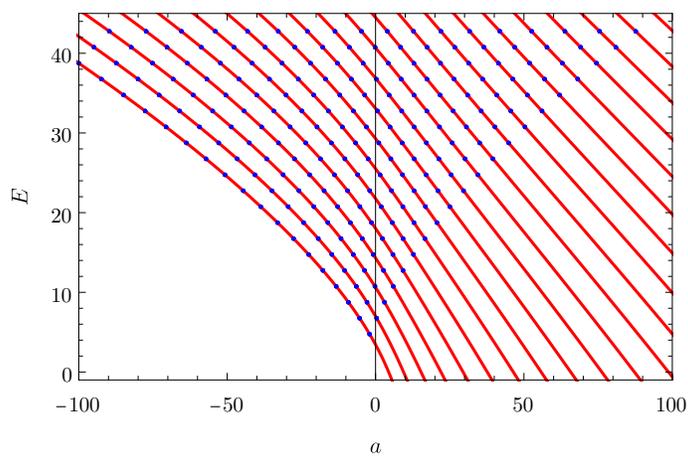}
\end{center}
\caption{Eigenvalues $E_{\nu,1}(a,1)$ for the perturbed Coulomb model
calculated by means of the truncation method (blue points) and a variational
method (red lines)}
\label{fig:CLHGQB1}
\end{figure}

Figure~\ref{fig:CLHGQB1} shows some eigenvalues for $\gamma =1$, $b=1$ in a
range of values of $a$ calculated by the truncation condition (\ref
{eq:trunc_cond_2}) and a variational method with the nonorthogonal basis set
of functions (\ref{eq:phi_CLH}).

\section{Conclusions}

\label{sec:conclusions}

Although there have been several excellent papers published on the subject
of conditionally solvable models\cite
{F79,FD80,M81,F81,F81b,RV91,FW81,D88,K89,BV90,CM95,BCD17} we have decided to
write the present one because the meaning of the exact solutions obtained
for such particular models have not been understood\cite
{V91,MHV92,FDMBB94,BM12,BB12,B14,B14b,BB14,FB15,BF15, FB12,BB13b,HMM20}.
These authors believe that the exact solutions to conditionally solvable
models are the only bound states supported by them and, as a consequence,
draw wrong conjectures such as the existence of allowed cyclotron
frequencies, allowed field intensities and the like. These wrong conclusions
stem from the fact that the exact solutions (with polynomial factors) are
possible for some particular values of the model parameters. The dependence
of the model parameters on the truncation number $n$ (the degree of the
polynomial factor) has been interpreted as the dependence of the parameters
on the quantum numbers and thereby the conclusion that bound states exist
only for particular value of certain experimental quantities. This wrong
interpretation of the truncation method has led them to believe that they
obtained the whole spectrum of a given model when they obtained just one (in
our second example) or a few (in our first example) energy for a given
model. We hope to have made this point clear in the present paper.

The paper of Child et al\cite{CDW00} reveals a most clear picture of the
distribution of the eigenvalues of the central-field sextic anharmonic
oscillator, as well as a hidden symmetry. In this paper we add somewhat
different diagrams of the distribution of the eigenvalues of conditionally
solvable models that we believe to provide additional valuable information.

\section*{Acknowledgements}

The research of P.A. was supported by Sistema Nacional de Investigadores
(M\'exico).

\appendix

\numberwithin{equation}{section}

\section{Symmetric tridiagonal matrix}

\label{sec:Symmetric matrix}

In this appendix we review a most interesting result derived by Child et al%
\cite{CDW00}. The three-term recurrence relations discussed above can be
rewritten as
\begin{equation}
U_{j}c_{j-1}+\left( V_{j}-\lambda \right) c_{j}+W_{j}c_{j+1}=0
\end{equation}
where $\lambda =E$ in the first example and $\lambda =a$ in the second one.
If we define new coefficients $\tilde{c}_{j}$ by means of the transformation
$c_{j}=Q_{j}\tilde{c}_{j}$ then we obtain the new eigenvalue equation
\begin{eqnarray}
&&M_{j,j-1}\tilde{c}_{j-1}+\left( M_{j,j}-\lambda \right) \tilde{c}%
_{j}+M_{j+1}\tilde{c}_{j+1}=0  \nonumber \\
&&M_{j,j-1}=U_{j}\frac{Q_{j-1}}{Q_{j}},\,M_{j,j}=V_{j},\,M_{j,j+1}=W_{j}%
\frac{Q_{j+1}}{Q_{j}}
\end{eqnarray}
If we set $c_{0}=\tilde{c}_{0}=1$, then $Q_{0}=1$.

If the matrix $\mathbf{M}$ is symmetric then its eigenvalues $\lambda $ are
real; therefore, we require that
\begin{equation}
M_{j+1,j}=U_{j+1}\frac{Q_{j}}{Q_{j+1}}=M_{j,j+1}=W_{j}\frac{Q_{j+1}}{Q_{j}}
\end{equation}
that leads to
\begin{equation}
Q_{j+1}^{2}=\frac{U_{j+1}}{W_{j}}Q_{j}^{2},\,j=0,1,\ldots
\end{equation}
Therefore, this matrix symmetrization is possible if $U_{j+1}/W_{j}>0$ for
all $j$ and, consequently, this condition is sufficient for the existence of
real eigenvalues $\lambda $. In the two examples discussed above $W_{j}$ is
always positive while $U_{j}$ becomes negative for a sufficiently great
value of $j$ unless we choose either $a$, $b$ or $E$ so that $U_{n+1}=0$.
This is exactly the truncation condition discussed in the preceding
sections. In other words: the truncation condition assures real eigenvalues $%
\lambda $.

\end{document}